\documentclass{article}
\usepackage{xcolor}
\usepackage[color=blue!50!green!70,draft]{todonotes} 
\usepackage{hyperref}
\usepackage{spconf,amsmath,graphicx}
\usepackage{bm}
 \usepackage{amsfonts}
 \usepackage{algpseudocode,algorithm,algorithmicx}
\algrenewcommand\algorithmicrequire{\textbf{Input:}}  

\DeclareMathOperator*{\argmin}{argmin}

\newcommand{\mux}{\mu_x}
\newcommand{\muy}{\mu_y}
\newcommand{\muz}{\mu_z}

\title{Dense Super-Resolution Imaging of Molecular Orientation via\\ Joint Sparse Basis Deconvolution and Spatial Pooling}

\name{Hesam Mazidi, Eshan S. King, Oumeng Zhang, Arye Nehorai,  Matthew D.  Lew\thanks{This work is supported by the National Science Foundation under grant number ECCS-1653777 and by the National Institute of General Medical Sciences of the National Institutes of Health under grant number R35GM124858 to M.D.L.}}
\address{Department of Electrical and Systems Engineering, Washington University in St. Louis, MO, USA}
\begin{document}
%
\maketitle
\begin{abstract}
In single-molecule super-resolution microscopy, engineered point-spread functions (PSFs) are designed to efficiently encode new molecular properties, such as 3D orientation, into complex spatial features captured by a camera. To fully benefit from their optimality, algorithms must estimate multi-dimensional parameters such as molecular position and orientation in the presence of PSF overlap and model-experiment mismatches. Here, we present a novel joint sparse deconvolution algorithm based on the decomposition of fluorescence images into six basis images that characterize molecular orientation. The proposed algorithm exploits a group-sparsity structure across these basis images and applies a pooling strategy on corresponding spatial features for robust simultaneous estimates of the number, brightness, 2D position, and 3D orientation of fluorescent molecules. We demonstrate this method by imaging DNA transiently labeled with the intercalating dye YOYO-1. Imaging the position and orientation of each molecule reveals orientational order and disorder within DNA with nanoscale spatial precision.
\end{abstract}

\section{Introduction}
\label{sec:intro}

Single-molecule localization microscopy (SMLM) relies on accurately and precisely estimating the position of many single emitters repeatedly blinking over time \cite{betzig2015Angewandte,hell2015Angewandte,moerner2015Angewandte}. 
Besides their position, fluorescent molecules interacting with their nano-environment also report information such as orientation and emission spectra that are hidden from standard, diffraction-limited optical imaging systems. Augmenting the PSF, or the impulse response of the  microscope, to uncover such fine details remains a vibrant research topic \cite{backlund2014ChemPhysChem,zhang2018apl}. In addition, biophysical measurements utilizing SMLM are more accurate when a physically-realistic image-formation model and recovery problem are utilized that include parameters such as molecular orientation and rotational mobility.

However, estimating multidimensional parameters such as molecular position, orientation, and brightness from noisy camera images poses a formidable challenge. For example, engineered PSFs have much larger footprints compared to the standard PSF (Fig. \ref{fig:figure1}(b)), thereby causing frequent PSF overlaps \cite{mazidi2018scireports} and lower pixel-wise signal-to-noise ratios, which can substantially reduce the detection rate of standard spot-detection methods.  Further, a target structure must be densely sampled by blinking emitters in order to be resolved, and therefore, it is vital to develop an algorithmic framework to address the aforementioned challenges. Joint estimation of the molecular position and orientation for the standard PSF has been considered previously \cite{patra2004JACS,aguet2009opticsexpress,mortensen2010naturemethods}. These methods, however, cannot be adapted to engineered PSFs for measuring jointly the position, 3D orientation, and rotational dynamics of fluorescent molecules. In addition, the problem of PSF overlap in the presence of multi-channel measurement errors has not been considered before. In this work, we tackle these challenges by 1) constructing a forward model that decomposes molecular parameters (i.e., brightness, position, and orientation) into six basis images corresponding to the six second moments of orientation dynamics \cite{zhang2018apl}; 2) building a   novel sparse basis deconvolution algorithm that exploits a group-sparsity norm to jointly recover all molecular parameters; and 3) applying spatial pooling across the six basis images to avoid false localizations due to experimental mismatches. We applied the proposed method to resolve DNA conformation via super-resolution imaging of DNA intercalating probes. 


\section{Image-formation model}
\label{sec:format}
\begin{figure}[t]
    \centering
    \includegraphics[width=1\linewidth]{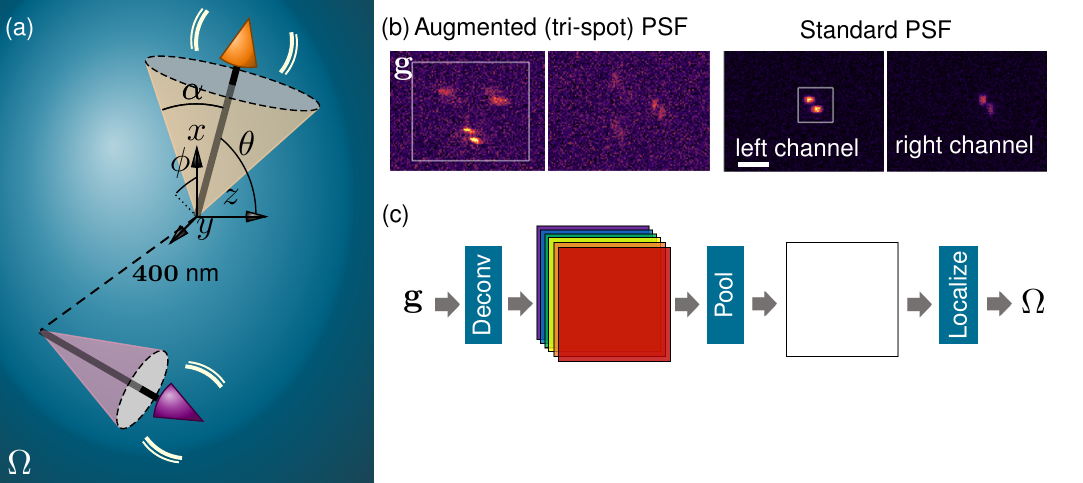}
    \caption{(a) Two closely-spaced dipole emitters with distinct orientations. (b) Images of emitters in (a) for (left) a microscope augmented with the Tri-spot PSF \cite{zhang2018apl} to produce orientation-sensitive images and (right) a standard microscope. White lines separate x- and y-polarized images. (c)~Proposed algorithmic framework consists of three stages.  Scale bar: 1 $\mu$m.}
    \label{fig:figure1}
\end{figure}

A dipole, such as a single fluorescent molecule, is characterized by its position $\bm{r}=[x,y]^T$  and an orientation vector $\bm{\mu}=[ \mux,\muy,\muz]^{T}=[\sin(\theta) \cos(\phi), \sin(\theta)\allowbreak \sin(\phi), \cos(\theta)]^{T}\in \mathbb{R}^{3}$, where $\theta$ and $\phi$ are the polar and the azimuthal angles in  spherical coordinates, respectively (Fig. \ref{fig:figure1}(a)). The  image of a dipole, denoted by $I$, can be described in terms of an orientational second-moment vector $\bm{\mathcal{M}}=[\langle\mux^2\rangle,\langle\muy^2\rangle,\langle\muz^2\rangle,\langle\mux \muy\rangle,\allowbreak \langle\mux \muz\rangle,\allowbreak \langle\muy \muz\rangle]^T$, in which $\langle \cdot \rangle$ represents the temporal average over a camera frame or equivalently an ensemble average over the orientation domain \cite{aguet2009opticsexpress,backer2014physchem}. More precisely, $I=s \Big(\sum_{j=1}^{6} \mathcal{M}^j \mathcal{B}^j\Big)$. Here, $s$ denotes the total number of photons emitted by the dipole, and the $\mathcal{B}^{j}$'s represent the so-called basis images  corresponding to a dipole exhibiting each  orientational second-moment component.  Note that here $I$ corresponds to two orthogonally-polarized images (i.e., x and y), concatenated together (Figs. \ref{fig:figure1}(b), \ref{fig:figure2}(c,d), \ref{fig:figure3}(a inset)). 

Consider the object domain $\Omega=\{\omega_i=(s_i, \bm{\mathcal{M}}_i, \bm{r}_i)\ |\ s_i \allowbreak \geq0, \Vert\bm{r}_i\Vert_\infty<r_{\text{max}}, i\in \{1,2,\ldots,P\}\}$ as a collection of $P$ dipole-like molecules located within a square region of interest of length 2$r_{\text{max}}$.  
We represent the PSF of  the microscope as $h(\bm{u};\omega): \Omega \rightarrow \mathcal{I}\times \mathbb{R}$ for which $\bm{u}$ denotes the coordinates in image space $\mathcal{I}\subset \mathbb{R}^{2}$. 
The noiseless image formed on the camera is the result of a weighted-convolution operation  $I(\bm{u})=\sum_{i=1}^{P} s_i h(\bm{u};\omega_i)*\delta(\bm{r}-\bm{r}_i)$ ($*$ denotes the (linear) convolution operator and $\delta(\cdot)$ represents the Dirac delta function).
We now exploit the fact that $h$ can be represented as the sum of six basis images:

\begin{align}\label{brightness-scaled-img-model}
        I(\bm{u})&=\sum_{i=1}^{P} s_i \Big( \sum_{j=1}^{6}\mathcal{M}^{j}_i  h^j(\bm{u}-r_i)\Big)
\end{align}
where $h^j(\cdot)$ corresponds to the basis image $\mathcal{B}^j$, and $\mathcal{M}_{i}^{j}$ represents the $j_{\text{th}}$ component of the second-order orientational moment of the $i_\text{th}$ emitter. As $s$  plays a scaling role here, we absorb it into $\bm{\mathcal{M}}$ and denote it by $\bm{\eta}=s \bm{\mathcal{M}}$. In the next section, we extend our model in Eq. (\ref{brightness-scaled-img-model}) to explicitly take into account the continuous position of molecules, which proves to be crucial for the robustness of the recovery algorithm.


Let $D$
be a set of $N$ discrete grid points, where the distance between two adjacent points is given by $2\rho$. 
A position vector  $\bm{r}$
can be uniquely mapped into $\bm{r}=\bm{d}+\bm{\delta} $ for some $\bm{d}\in D$ and $\bm{\delta}=[\delta_x,\delta_y]^{T}$ with $\delta_x, \delta_y  \in [-\rho,\rho)$. Therefore, we can rewrite Eq. (\ref{brightness-scaled-img-model}) as $I(\bm{u})= \sum_{i=1}^{N} \sum_{j=1}^{6} \eta^{j}_i h^j(\bm{u}-\bm{d_i-\delta_i})$. Note that $\bm{\eta}_i=0$ if no molecule can be  mapped to a point in the neighbourhood of $\bm{d}_i$. For a fixed $i\in \{1,2,\ldots,N\}$ and $j\in\{1,2,\ldots,6\}$, we approximate $h$ via first-order Taylor expansion $h^{j}(\bm{u-d_i-\delta_i})\approx h^{j}(\bm{u}-\bm{d_i})-h'^{j}(\bm{u-d_i})^{T} \bm{\delta}_i$, in which $h'(\cdot)$ denotes the derivative of $h$. Within the first-order approximation, we have
\begingroup
\allowdisplaybreaks
\begin{align}\label{first-order-approx}
I(\bm{u})=&\sum_{i=1}^{N} \sum_{j=1}^{6} \Big(\eta^{j}_i h^j(\bm{u}-\bm{d_i})-  {h'_x}^{j}(\bm{u}-\bm{d_i}) \zeta_{x,i}^{j}-\nonumber\\ 
&{h'_y}^{j}(\bm{u}-\bm{d_i}) \zeta_{y,i}^{j}\Big),
\end{align}
\endgroup
where $\bm{\zeta}_{i}^{j}= \eta^{j}_i \bm{\delta}_i=[\zeta_{x,i}^{j},\zeta_{y,i}^{j}]^T$; ${h'_x}^{j}$ and ${h'_y}^{j}$ represent derivatives of $h$ along the $x$-axis and $y$-axis, respectively. Remarkably,  the image model in  Eq. (\ref{first-order-approx}) can be interpreted as the sum of a few PSFs weighted by object-domain parameters $\Omega$. 

We also take into account camera pixelation by integrating the image $I$ over $m$ pixels, which effectively results in  matrices $\bm{\Phi}^{j}, \bm{Gx}^{j}$ and $\bm{Gy}^{j}$ for each basis with index $j$. Note that $\bm{\Phi}$ corresponds to $h$ in Eq. (\ref{first-order-approx}) whereas $\bm{Gx}$ and $\bm{Gy}$ correspond to $h'_x$ and $h'_y$, respectively. The final imaging model can be compactly represented as $\mathcal{A} \bm{F} =\sum_{j=1}^{6}\bm{A}^j \bm{f}^j$ with  $\bm{A}^j=[\bm{\Phi}^{jT},\bm{Gx}^{jT},\bm{Gy}^{jT}]^T$ and $\bm{f}^{j}=[\bm{\eta}^{jT},\bm{\zeta}_x^{jT},\allowbreak \bm{\zeta}_y^{jT}]^T$. We further model the photon count in each pixel $i$ as an independent Poisson distribution $g_i \sim \text{Pois}\Big((\mathcal{A}\bm{F}+\bm{b})_i\Big)$ 
whose mean equals the sum of detected photons emitted by fluorescent  molecules (i.e., $\mathcal{A}\bm{F}$) and  background flux (i.e., $\bm{b}$). 

Therefore, the nonlinear image-formation model for dipole emitters parameterized by brightness, 3D orientation, rotational dynamics, and position can be represented by a linear model utilizing a set of new parameters, accompanied by nonlinear constraints.
\section{Recovery algorithm}
Next, we focus on the regularized inverse problem for the derived model and show how we can handle nonlinear constraints to obtain efficient recovery in the presence of image overlap and experimental mismatches. The proposed algorithm consists of three main stages (Fig. \ref{fig:figure1}(c)) as follows.
\subsection{Joint sparse basis deconvolution}
Our main observation is that the number of underlying molecules is much smaller than the number of ambient parameters.
Taking advantage of the cascading structure of the proposed signal model, this prior knowledge can be encoded via  a group/joint-sparsity norm across the six basis images,  hence the name joint sparse basis deconvolution.

We formulate the recovery problem as a minimization:

\begin{equation}\label{joint-sparse-basis-deconv}
\hat{\bm{F}}=\argmin_{\bm{F}} \{\mathcal{L}(\bm{F})+\lambda \mathrm{R}(\bm{F})+\mathcal{I}_{\mathcal{C}}(\bm{F})\},    
\end{equation}
where $\mathcal{L}$ is the negative Poisson log-likelihood; $\mathrm{R}$ denotes the regularizer (described below); $\lambda$ is a penalty parameter; and $\mathcal{I}_{\mathcal{C}}$ denotes the  indicator function of the set $\mathcal{C}$ representing the constraint set.
The regularizer, $\mathrm{R}$, is a group-sparsity norm for brightness and position gradients across all six basis images defined by $\mathrm{R}(\bm{F})=\sum_{i=1}^{N} \sqrt{\sum_{j=1}^{6}\Big((\eta^j_i)^2+(\zeta^j_{x,i})^2+ (\zeta^j_{y,i})^2} \allowbreak\Big)$.
The set $\mathcal{C}=\cap_{i=1}^{6}C_i$ captures  physical constraints on the signal model. In particular,  $C_j=\{\bm{f}^j\ |\ \bm{\eta}^j\geq 0,\  -\rho\bm{\eta}^j\leq \bm{\zeta^j_x}\leq \rho\bm{\eta}^j,\ -\rho\bm{\eta}^j\leq \bm{\zeta^j_y}\leq \rho\bm{\eta}^j\}$ for $j=\{1,2,3\};\ C_j=\{\bm{f}^j\ |\ \Vert \bm{\zeta^j_x}\Vert_\infty \leq \rho \Vert \bm{\eta^j}\Vert_\infty,\ \Vert \bm{\zeta^j_y}\Vert_\infty \leq \rho \Vert \bm{\eta^j}\Vert_\infty \}$ for $j=\{4,5,6\}$. One may note that the sets $C_j$ ($j=\{4,5,6\}$) are non-convex. To preserve convexity, we neglect the first-order approximation in the last three basis images, implying that $\bm{f}^{j}=[\bm{\eta}^{jT}, \bm{0}^T, \bm{0}^T]^T$ for $j=\{4,5,6\}$. Henceforth we drop the corresponding constraint sets.  This approximation is physically justified since the  energy radiated by a dipole is mostly contained in the first three basis images \cite{zhang2018apl,backer2014physchem}. 
Consequently, it renders Eq. (\ref{joint-sparse-basis-deconv}) as a convex program, and we develop a variant of FISTA to solve it \cite{mazidi2018scireports,beck2009SIAM}.

Deriving a closed-form proximal operator required for FISTA in Eq. (\ref{joint-sparse-basis-deconv}) is not possible. We thus smooth the regularizer term with a differentiable function, e.g., its Moreau
 envelope. Let $w(\bm{F})=\lambda \mathrm{R}(\bm{F})$. The Moreau envelope of $w$, $\mathcal{E}_\tau(w)$, is continuous, and its gradient is given as $\nabla w_\tau (\bm{F})=\frac{1}{\tau}(\bm{F}-\text{prox}_{\tau w}(\bm{F}))$ in which $\text{prox}_{\tau w}(\bm{F})$ is the proximal operator of $\tau \lambda \mathrm{R}(\bm{F})$. The smoothing parameter $\tau$ controls the accuracy of the approximation. The modified optimization  is now an instance of the gradient projection given as $\min_{\bm{F}} \mathcal{L}(\bm{F})+\mathcal{E}_\tau(\lambda \mathrm{R}(\bm{F}))+ \mathcal{I}_{\mathcal{C}}(\bm{F})$,
 for which $\mathcal{I}_{\mathcal{C}}(\cdot)$ is the indicator of the set ${\mathcal{C}}=\cap_{i=1}^{3}C_i$. Another difficulty is that the projection operator associated with ${\mathcal{C}}$ does not admit a closed-form expression. To tackle this issue, we approximate each $C_{j}$ with a \textit{second-order cone} as $\widehat{C}_{j}=\{\bm{f}^j\ |\ \Vert[\zeta^j_{x,i},\zeta^j_{y,i}]^{T} \Vert_{2} \leq  \rho \eta_i^{j},\ i=\{1,2,\ldots,N\}\}$ ($j=\{1,2,3\}$), which quite remarkably admits a closed-form projection operator \cite{mazidi2018scireports}. 
\subsection{Spatial pooling for robust emitter identification and localization}
An important challenge in sparse recovery is that of model mismatch, which has been shown to degrade the performance of grid-based recovery algorithms. However, for the problem of emitter localization,  a model mismatch can cause false emitter localization, thereby introducing \textit{bias} in the measurements (Fig. \ref{fig:figure2}(a)). In our imaging system, the misalignment of two polarization channels leads to a mismatch; the estimated positions obtained separately from left and right channels may differ by an amount comparable to the localization precision. Notice, however, that the \textit{average} of position estimates in two channels can actually be considered as the main parameter of interest.
Using this insight and the fact that our model can provide \textit{continuous} position estimates, we next show how to robustly identify the correct number of molecules.

As shown in \cite{mazidi2018scireports}, the recovered joint signal $\hat{\bm{F}}$ exhibits a specific structure in which position gradients converge to the true location of each emitter. To exploit this structure, we construct an operator called GradMap $\mathcal{G}: \mathbb{R}^{3N} \rightarrow \mathbb{R}^{N}$. Briefly, for each grid point or pixel, $\mathcal{G}$ computes how much the position gradients in neighbouring points converge to the grid point of interest. Put differently, $\mathcal{G}$ returns the ``likelihood'' that a true emitter belongs to each pixel.  We apply $\mathcal{G}$ on each of first three deconvolved basis images to obtain $G^j=\mathcal{G}(\bm{\hat{f}}^j)$ for  $\bm{\hat{f}}^j=[\bm{\hat{\eta}}^{jT},\bm{\hat{\zeta}}_x^{jT},\allowbreak \bm{\hat{\zeta}}_y^{jT}]^T$ and $j=\{1,2,3\}$ (Fig. \ref{fig:figure2}(b)). The final position estimates are obtained by pooling these three spatial maps, i.e., $G=\frac{1}{3}\sum_{i=1}^{3} G^i$ (Fig. \ref{fig:figure2}(b)). The emitters' initial positions and brightnesses correspond to the local maxima of $G$.

Let  $\text{Supp}(\hat{\bm{F}})$ be the support set of estimated emitters' positions obtained via spatial pooling. Localization is achieved via solving a constrained maximum-likelihood problem with an initial point obtained in the previous stage:
\begin{align}
    \min_{\bm{F} \in \mathcal{C}\cap \text{Supp}(\hat{\bm{F}}) } \mathcal{L}(\bm{F}),
\end{align}
which is a convex program. The molecular orientation parameters are estimated using the recovered second-moment vectors $\hat{\bm{\eta}}=\hat{s}\widehat{\bm{\mathcal{M}}}$. 

\section{Results}
\begin{figure}[t]
    \centering
    \includegraphics[width=1\linewidth]{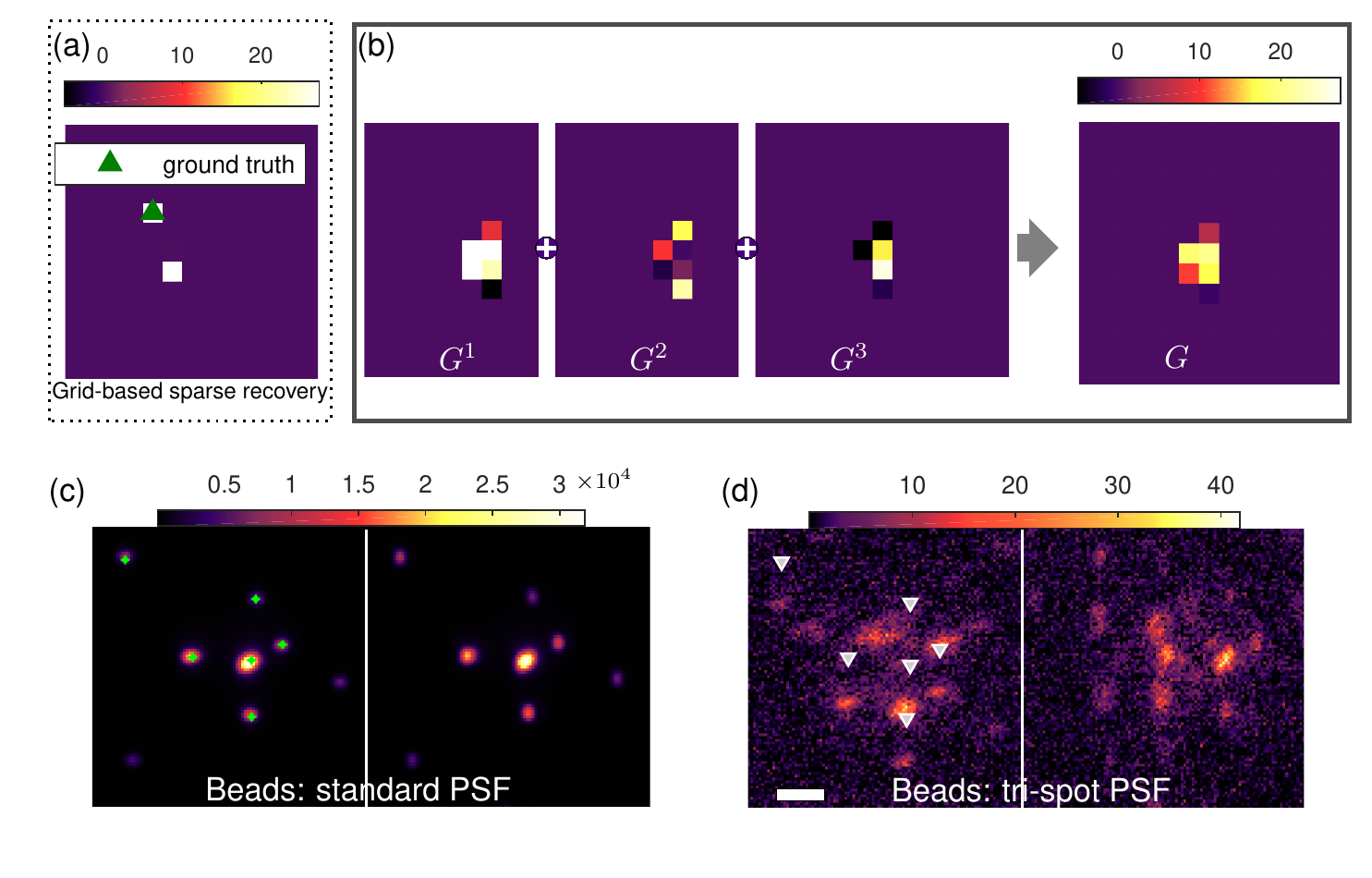}
    \caption{(a) False discovery of a single emitter using a grid-based sparse recovery algorithm due to a channel misalignment of one camera pixel. Accurate counting of emitters is restored via a pooling strategy. (b) (left) Recovered GradMaps for the first three bases and (right) the result of pooling these maps. (c) Experimental images of beads using standard PSF (100 ms exposure). (d) Images of beads in (c) using the Tri-spot PSF and artificially reduced SNR (5~ms exposure). The green dots in (c) and gray triangles in (d) display the localizations recovered by the proposed algorithm. Scale bar: 1~$\mu$m. Color bars: (a,c,d) photons and (b) scaled-photon likelihood per $58\times 58\ \text{nm}^2$ pixel.}
    \label{fig:figure2}
\end{figure}
Fluorescent beads were utilized to validate our method's detection and localization capabilities. A sparse layer of beads 
was imaged using the standard and Tri-spot PSFs with the system described in \cite{zhang2018apl} (Fig. \ref{fig:figure2}). 
Despite the substantial overlap and channel mismatch, as evidenced by the offset of the green dots from the peaks of the standard PSF (Fig. \ref{fig:figure2}(c)), the algorithm is able to recover the correct position and number of emitters using the Tri-spot PSF (Fig. \ref{fig:figure2}(d)). 

\begin{figure}[t]
    \centering
    \includegraphics[width=1\linewidth]{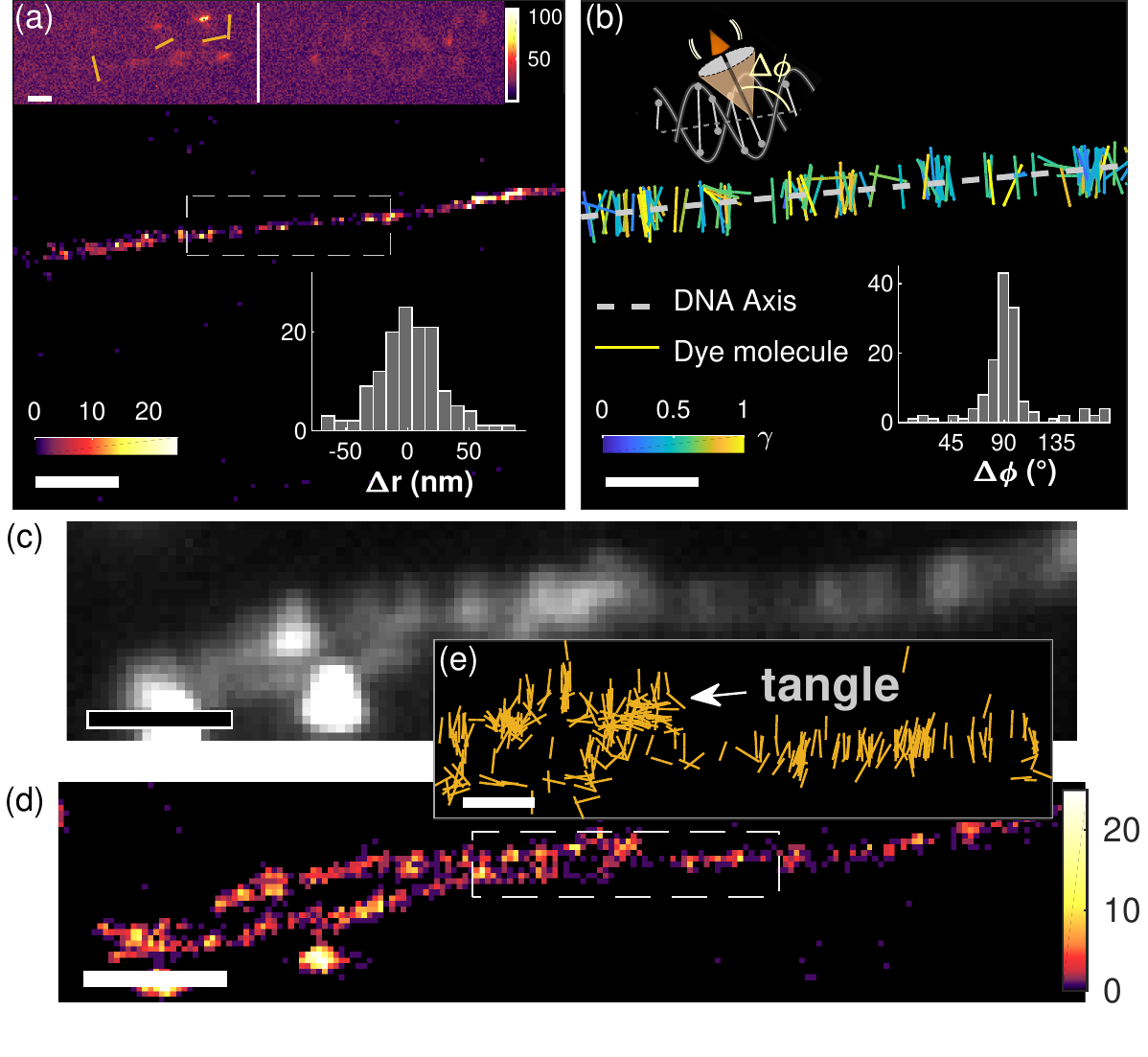}
    \caption{(a) Super-resolved image of $\lambda$-DNA. Inset upper-left: a raw image showing the recovered molecular orientations from four detected PSFs. Inset lower-right: distribution of localization distances from the DNA axis ($\sigma_{\Delta r}=26$~nm). (b)~Dye orientations within the region denoted by a gray dashed rectangle in (a), overlaid on the estimated DNA axis (white dashed line) and color-coded by rotational constraint $\gamma$.
    Inset upper-left: dye intercalation into a DNA helix. Inset lower-right: distribution of in-plane dye orientation relative to DNA axis ($\Delta \phi_{avg} = 94^{\circ}$, $\sigma _{\Delta \phi}=28^{\circ}$). (c) Diffraction-limited image of a DNA strand. (d) Super-resolved image corresponding to (c). (e) Dye orientations reveal order and disorder (i.e. tangles) within the region denoted by the dashed gray rectangle in (d). Scale bars: (a,c,d) 1 $\mu$m, (b) 400 nm, (e) 200 nm. Color bars: (a,d) localizations and (a inset) photons per $58\times 58\ \text{nm}^2$ pixel.}
    \label{fig:figure3}
\end{figure}

$\lambda$-DNA (Thermo Scientific) was deposited onto coverslips using a molecular combing technique \cite{Backer2016} and imaged using a reducing-oxidizing buffer \cite{BALM} and Tri-spot PSF. 
The proposed method recovered the location, orientation, and rotational mobility $\gamma$ ($\gamma$ = 0 for an isotropic emitter and $\gamma$ = 1 for a fixed dipole) of many blinking YOYO-1 dyes transiently bound to DNA (Fig. \ref{fig:figure3}(a,b)). The long-axis of the DNA strand was estimated by a simple least-squares polynomial fit, and the orientation of the dyes bound to the strand ($\Delta \phi$) was calculated relative to this fit. As shown in Fig. \ref{fig:figure3}(b), molecules detected along a linear section of DNA are mostly oriented perpendicular to the DNA axis, which is consistent with the primary binding mode of YOYO-1 \cite{YOYOBinding,cruz2016quantitative}. The distribution of binding angles is likely broadened by secondary dye binding modes along the major and minor grooves of DNA, i.e., along its axis \cite{YOYOBinding}. These modes have been reported at high concentrations of YOYO-1 and can be observed in regions of high localization density (Fig. \ref{fig:figure3}(b)), where some dipoles are oriented parallel to the strand. The recovered rotational constraint yielded an average ``wobble'' cone angle of 91$^{\circ}$, assuming uniform rotation within a cone. Our 3D estimate of wobble angle is larger than previously observed for YOYO-1 \cite{cruz2016quantitative} and SYTOX Orange \cite{Backer2016} (another DNA intercalating dye), which were 2D orientation measurements. As YOYO-1 is a bis-intercalator, each molecule is effectively two dipole emitters with similar in-plane orientations but offset out-of-plane orientations; this superposition of dipoles likely causes a larger effective wobble angle in 3D and therefore an overestimate of the true rotational mobility of each bis-intercalator. Previous 2D studies of YOYO-1 were not sensitive to this out-of-plane offset.

The algorithm was also able to distinguish areas of relative organization and disorder. Regions where the average dye orientation was not perpendicular to the DNA axis indicate local fluctuations, or tangles, in the DNA strand (Fig. \ref{fig:figure3}(e)) that are not observable via standard SMLM (Fig. \ref{fig:figure3}(d)). 

\section{Conclusion}
Engineered PSFs for orientation-sensitive super-resolution imaging pose major challenges, such as frequent overlapping PSFs and channel registration errors, for standard localization algorithms. We have presented a novel, robust algorithm for simultaneous recovery of the position and 3D orientation of fluorescent molecules using engineered PSFs. In contrast to methods based on defocus imaging, our algorithm can be applied to arbitrary orientation-sensitive PSFs and remedies mislocalizations due to PSF overlap and channel misalignment. We validated this method by imaging $\lambda$-DNA labeled with an intercalating dye, showing that the recovered molecular dipole orientation is primarily perpendicular to the DNA axis, which is consistent with previous observations. Interestingly, by measuring the full 3D orientation of YOYO-1, we observe a rotational constraint that is significantly smaller than that measured by 2D methods, suggesting that 3D orientation measurements may be necessary for revealing the true rotational dynamics of single molecules.
\bibliographystyle{IEEEbib}
\bibliography{strings,refs}

\end{document}